\documentclass[aps,prl,twocolumn,preprintnumbers,showpacs,nofootinbib]{revtex4}
\usepackage{graphicx}
\usepackage{dcolumn}
\usepackage{bm}
\newcommand{\beq}{\begin{equation}}
\newcommand{\eeq}{\end{equation}}
\newcommand{\beqa}{\begin{eqnarray}}
\newcommand{\eeqa}{\end{eqnarray}}

\newcommand{\eeqn}{\end{equation}}

\newcommand{\eeqan}{\end{Eqnarray}}

\renewcommand{\bar}[1]{\overline{#1}}

\newcommand{\lsim}{\mathrel{\raise.3ex\hbox{$<$\kern-.75em\lower1ex\hbox{$\sim$}}}}
\newcommand{\gsim}{\mathrel{\raise.3ex\hbox{$>$\kern-.75em\lower1ex\hbox{$\sim$}}}}

\newcommand{\mw}{m_W}
\newcommand{\mt}{m_t}

\newcommand{\msb}{{\bar{\scriptscriptstyle M \kern -1pt S}}}

\newcommand{\herw}{{\small HERWIG}}

\newcommand \etmiss     {\ifmmode E_{\sss T}^{\sss miss} \else
                            $E_{\sss T}^{\sss miss}$\fi}
\newcommand \ptl        {\ifmmode p_{\sss T}^{\sss \ell} \else
                            $p_{\sss T}^{\sss\ell}$  \fi}

\newcommand     \sss            {\scriptscriptstyle}

\newcommand{\anfis}[1]{Ann. Phys. #1}

\def        \ATLAS      {\mbox{ATLAS}}

\def        \infb       {\mbox{fb$^{-1}$}}

\def        \mjjb       {\mbox{$m_{jjb}$}}

\def        \mW         {\mbox{$m_W$}}

\newcommand \avmlb {\ifmmode \langle m_{\ell b}^{\sss 2}\rangle \else
  $\langle m_{\ell
    b}^{\sss 2} \rangle $\fi}
\newcommand \avthetalb {\ifmmode
  \langle \cos\theta_{\ell b}\rangle \else $\langle \cos\theta_{\ell
  b}\rangle $\fi}

\def\beq{\begin{equation}}
\def\eeq{\end{equation}}
\def\bea{\begin{eqnarray}}
\def\eea{\end{eqnarray}}
\newcommand{\be}{\begin{equation}}

\def\beqa{\begin{eqnarray}}
\def\eeqa{\end{eqnarray}}
\def\beq{\begin{equation}}
\def\eeq{\end{equation}}

\let\gam=w

\renewcommand{\epsilon}{\varepsilon}

\def \pt   {\mbox{$p_{\scriptscriptstyle T}$}}

\def \bbbar {\mbox{$b \bar b$}}

\def \pt   {\mbox{$p_{\scriptscriptstyle T}$}}

\def    \mt             {\mbox{$m_t$}}
\newcommand \jpsi{\ifmmode{J/\psi
    }\else{$J/\psi$}\fi}

\def        \infb       {\mbox{fb$^{-1}$}}

\def        \mjjb       {\mbox{$m_{jjb}$}}

\def        \mW         {\mbox{$m_W$}}

\begin{document}
\input epsf \renewcommand{\topfraction}{0.8}

\preprint{UCI-TR-2006-15}  \preprint{hep-ph/0609244}

\title{Searching for $CPT$ violation in $t\bar t$ production}

\author{J. A. R. Cembranos}
\affiliation{Department of Physics and Astronomy,
 University of California, Irvine, CA 92697 USA}
\author{A. Rajaraman}
\affiliation{Department of Physics and Astronomy,
 University of California, Irvine, CA 92697 USA}
\author{F. Takayama}
\affiliation{Institute for High-Energy Phenomenology, Cornell
University, Ithaca, NY 14853, USA}

\date{\today}

\begin{abstract}

We analyze the possibility of observing $CPT$ violation in the top
sector. We present current bounds on $CPT$ violation in this sector,
and analyze the prospects of improving these bounds at the LHC.

\end{abstract}


\maketitle

\section{Introduction}

The discrete symmetries of charge conjugation ($C$), parity ($P$) and
time reversal symmetry ($T$) were long believed to be exact
symmetries. However, as is well known, the weak interactions were
 found to violate $C$ and $P$ maximally \cite{P,PDG}. While the product $CP$ seemed
 to be conserved,
 evidence for $CP$-violation was later found in kaon decays \cite{CP,PDG}.
On the other hand, the triple product $CPT$ has never been seen to be
violated in any experiment, and
in fact, there are several important tests that constrain the
magnitude of $CPT$ violation in various sectors of the Standard
Model.

However, given our past experience with discrete symmetries, it is
important to actively search for potential $CPT$-violating effects.
Many well-motivated extensions of the standard model have such effects \cite{$CPT$models}.
String theory is inherently a nonlocal theory, and may well be a
source of  $CPT$ violation \cite{$CPT$string}. Noncommutative
theories \cite{Seiberg:1999vs} explicitly break Lorentz invariance, and hence could break
$CPT$. Models of ghost condensation \cite{Arkani-Hamed:2003uy} also spontaneously break
Lorentz invariance.

There are also some experimental motivations for considering models
with broken $CPT$. In particular, 
$CPT$ violation in the neutrino sector has been proposed as an
explanation for the combined LSND, atmospheric and solar oscillation
data \cite{neutrinos}.

Finally, we note that $CPT$ violation is an unambiguous signal of new
physics because this discrete symmetry is automatically conserved by
any local relativistic quantum field theory \cite{local}. Therefore
it is very interesting to check whether this is truly an exact
symmetry  of the Standard Model.

It is particularly interesting to search for $CPT$ violation in the
top quark sector. The top quark is the most massive known elementary particle, and the only fermion
with an unsuppressed coupling to the electroweak symmetry breaking
sector. This suggests that it may also probe $CPT$ violation to a
greater degree than the other fields of the Standard Model. There are
no current bounds on $CPT$ violation in this sector, and in this
paper, we will present the first such bounds, and analyze the best
direction for further improving them.

A difference in the masses and lifetimes of a particle and its
antiparticle is a model independent signature of $CPT$ violation. For
any particle $a$, with an antiparticle $\bar{a}$, we can parametrize
$CPT$ violation for that particle by  the dimensionless
quantity $R_{CPT}(a)\equiv 2(m_a-m_{\bar a})/(m_a+m_{\bar a})$.
Bounds on this quantity for various particles are given in Table
\ref{$CPT$}.

\begin{table}[ht]
\centering
\begin{tabular}{||c|c||}
\hline\hline
Particle (a)& $R_{CPT}(a)\equiv\frac{2(m_a-m_{\bar a})}{m_a+m_{\bar
a}}$
\\
\hline
$W^+$&$(-2\pm 7)\times10^{-3}$
\\
$e^+$&$<\,10^{-9}$ (90\% c.l.)
\\
$\pi^+$&$(2\pm 5)\times10^{-4}$
\\
$K^+$&$(-0.6\pm 1.8)\times10^{-4}$
\\
$K^0$&$<\,10^{-18}$ (90\% c.l.) *
\\
$p$&$<\,10^{-8}$ (90\% c.l.) *
\\
$n$&$(9\pm 5)\times10^{-5}$
\\
\hline\hline
\end{tabular}
\caption{\label{$CPT$} Experimental constraints on the  fractional
mass difference between particle ($a$) and antiparticle
 ($\bar{a}$) for various particles: $R_{CPT}(a)\equiv 2(m_a-m_{\bar a})/(m_a+m_{\bar a})$ .
 The * means that the constraint applies to the absolute value $|R_{CPT}|$ \cite{PDG}.}
\end{table}

In this paper we will focus on the measurement of
 $R_{CPT}(t)\equiv 2(m_t-m_{\bar t})/(m_t+m_{\bar t})$
\cite{Cembranos:2005ch,Juste:2006sv}.
In the following sections, we will describe various methods for
measuring a difference between $m_t$ and $m_{\bar t}$ in  top
anti-top  events at  colliders. We study the di-lepton channel, and
also
 the semileptonic channels, which appear to be
more promising. We find the first bound on $R_{CPT}(t)$ using
Tevatron data, and discuss future prospects for improving these
bounds at the LHC and ILC.

\section {Di-lepton channel at hadronic colliders}

Di-lepton events occur through the decay
 $t\bar t \rightarrow W^+
(\rightarrow l^+\nu)\,b\,W^-(\rightarrow l^-\bar \nu)\,\bar b $,
 with $l = e$ or $ \mu$. This decay mode  has been used at the Tevatron to
measure the top quark pole mass assuming an identical mass for the
top and anti-top quarks. If the masses are different, this can
potentially be seen in the di-lepton channel.

\begin{figure}[ht]
 \centerline{
    \includegraphics[width=0.4\textwidth,clip]{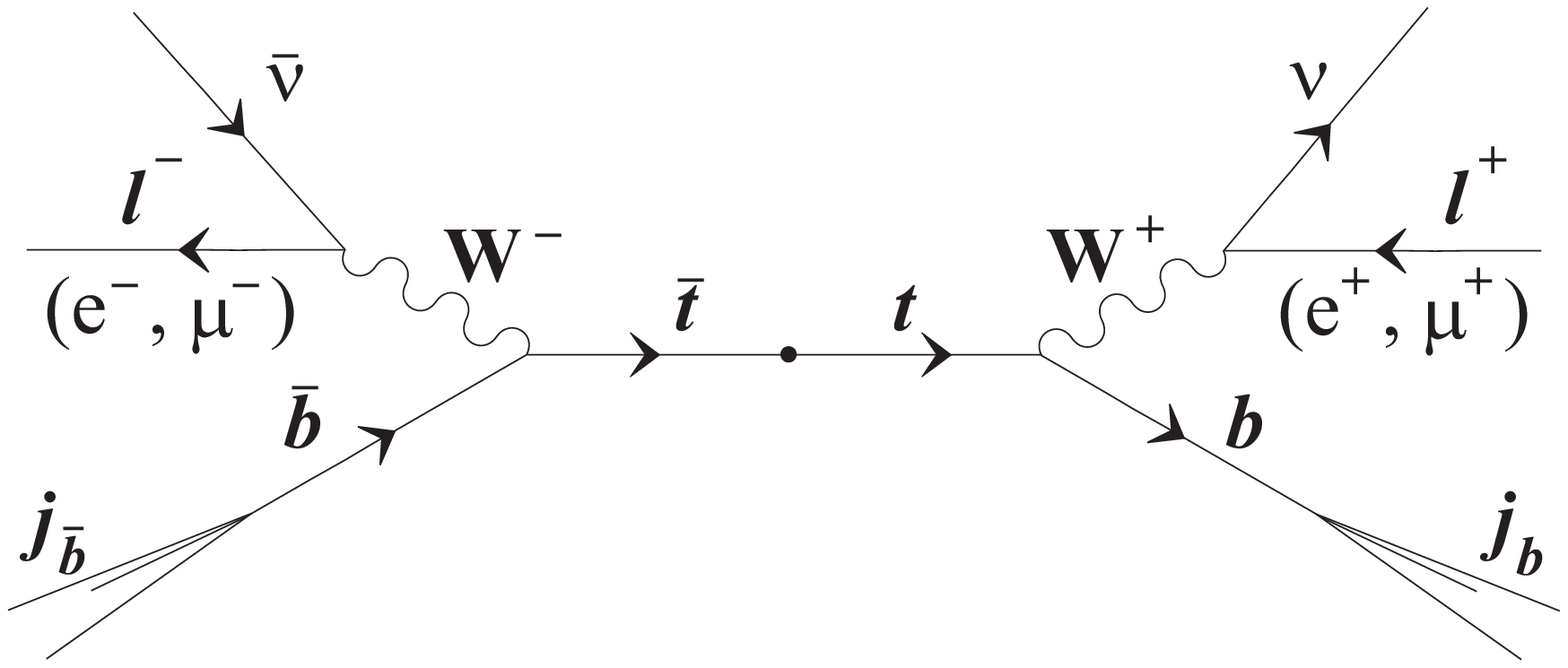}}
  \caption{Schematic example of the top and anti-top decays in the dilepton channel.}
  \label{dileptonSch}
\end{figure}

The signature of a di-lepton event consists of two isolated leptons
with high transverse momentum (\pt), high missing transverse energy
(\etmiss) due to the undetected neutrinos and two jets coming from
the $b$-quarks ($j_{\bar b}$ and $j_{b}$).
The background comes from \bbbar, $W^+W^-$+jets and $Z/\gamma(\rightarrow
l^+l^-)$+jets production, and can be reduced by appropriate cuts. For
instance,  CDF requires $m_{l^+l^-}\notin(75,105)$ GeV in order to exclude
 $Z \rightarrow l^+l^- X$ events \cite{CDFdilept1}.
After these cuts are imposed, the remnant background
is dominated by the Drell-Yan $Z/\gamma(\rightarrow l^+l^-)$ events.

Several indirect methods have been used to reconstruct the top mass using
di-lepton events. For instance, one technique takes advantage of the
correlation between
 the energy of the two leading jets (from the $b$ and $\bar b$ quarks) and the top
mass (or anti-top) \cite{CDFdilept1}.   However in this analysis, it
is implicitly assumed that the top and anti-top masses are the same,
and hence one will measure  the average value
 $\bar m_{t\bar t}=(m_{t}+m_{\bar t})/2$. It is difficult to
 obtain an estimate of a mass difference using this technique.

A better approach is to use the invariant mass of the lepton and $b$
quark coming from the single decay of the top (or the
$CPT$ conjugate process for the anti-top). 
The value of \mt~can then be estimated using the expression
\cite{CDFdilept1}:
\begin{equation}
  \mt^2 = \mw^2+2\avmlb/[1-\avthetalb]\,,
\label{mllb}
\end{equation}
where terms involving lepton and $b$ quark masses have been
neglected. Here \avmlb~is the squared mean invariant mass of the
lepton and $b$-jet. 
The mean value of \avthetalb, the angle between the lepton and the
$b$-jet in the $W$ rest frame, can be evaluated from the SM
tree-level calculation: $\avthetalb=\mw^2/(\mt^2+2\mw^2)$.

In general, it is not possible to determine which lepton ($l^+$ or
$l^-$) should be paired with each $b$-jet ($j_{\bar b}$~  or
$j_{b}$). A simple criterion, which provides a correct pairing close
to 75\% of the time, consists of selecting the pairing with smaller
value of \avmlb~ (the performance of this method is improved for the
LHC with respect to the Tevatron \cite{CDFdilept1,mtdetLHC}). So this
method could be sensitive to a difference between the top and
anti-top mass. Indeed the reconstructed mass distribution should show
two different peaks, and can be observed in collider experiments if
the mass difference is  large enough.

\begin{figure}[ht]
  \begin{center}
    \includegraphics[width=0.3\textwidth,clip]{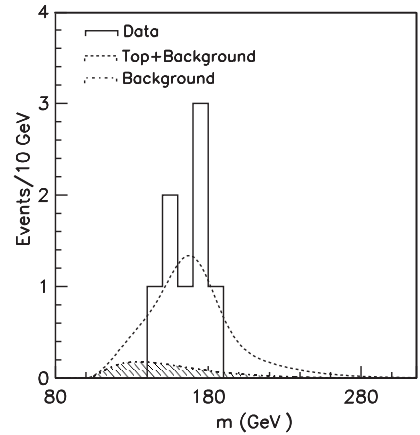}
  \end{center}
  \caption{Reconstruction of the top mass from the eight di-lepton events collected by
  the CDF Experiment (solid) in relation with the  1.3 expected background events distribution (shaded) and the additional 6.7  top Monte Carlo events (dashed). See \cite{CDFdilept2} for details. }
\label{CDFdilept}
\end{figure}

We can find a constraint on the mass difference between the top and
anti-top   using the data accumulated at CDF \cite{CDFdilept1}
from 1992 through 1995. (While there appear to be two peaks, the
uncertainties are large, and in any case, the  analysis in the
semileptonic jet channels rules out the possibility that this is due
to a mass difference.) The analysis using the above technique is
consistent with the same mass for top and anti-top, i.e. with only
one  distribution peaked at: $m_t=m_{\bar t}=163\pm 2 (stat.)\pm 9
(syst.)$ GeV.

We can estimate $R_{CPT}(t)$
adding directly the  uncertainties in order to be conservative.
At  95\% confidence level, we find $|R_{CPT}(t)|<0.13$.
This bound is dominated by the systematic errors, which come mainly
from the uncertainties in the jet energy scale and the shape of the
background distributions \cite{CDFdilept1}.

We can also estimate the sensitivity of future experiments to
$R_{CPT}(t)$. For  the LHC, with an integrated luminosity of
10~\infb, the expected statistical uncertainty on \mt~using this
technique is estimated to be $\pm0.9$~GeV. Systematic errors from
initial and final state radiation
together are about 1~GeV. An uncertainty of the jet energy scale of
1\% will
 produce a systematic error of about 0.6~GeV \cite{mtdetLHC}.
Therefore one can expect a sensitivity $|R_{CPT}(t)|\simeq 0.026$ at
the 95\% c.l. (supposing $\bar m_{t\bar t}\equiv (m_t-m_{\bar t})/2=
174.3$ GeV\cite{PDG}). Thus the LHC data could be sensitive to a
difference of order $4.5$ GeV between the top and anti-top masses.

Other techniques may also be used to measure the top mass. For
example, one can use the invariant mass distribution of the two
leptons ($m_{l^+l^-}$) in tri-lepton events \cite{mtdetLHC} to
measure the top mass. In fact, it is possible to obtain a more
precise measurement by using all available information in the event
\cite{CDFdilept2,D0dilept} (See Figure \ref{CDFdilept}).

The use of these techniques to measure the top mass could
improve the bound on $R_{CPT}(t)$. However the current bounds cannot
be translated to $R_{CPT}(t)$ in a straightforward way. The analysis
should take into account from the beginning the possibility of a
difference between the top and anti-top masses, eliminating the
constraint $m(l^+ \nu b)=m(l^- \bar \nu \bar b)$, which is usually
assumed in such reconstruction techniques \cite{CDFdilept2,D0dilept}.

\section {Lepton plus jets channel at hadronic colliders}
\label{sec:masslj} In any case, it is more promising to consider
semileptonic decays, where one of the $W$ decays into leptons and the
other  into hadrons: $t\bar t \rightarrow W^+(\rightarrow l^+\nu
b)\,b\,W^-(\rightarrow q \bar q')\,\bar b$.
The invariant mass of the three jets coming from the top (\mjjb
$\equiv m_{j_q j_{\bar q} j_{b}}$) or anti-top (\mjjb $\equiv m_{j_q
j_{\bar q} j_{\bar b}}$)
has a peak at the top (\mt) or anti-top ($m_{\bar t}$) mass
respectively.
The leptonic part of the decay can be used to tag the event, which is
characterized by an isolated lepton with a high \pt~and large \etmiss
coming from the undetected neutrino.

In addition proper cuts must be imposed to reduce the background,
which is given by the following processes: \bbbar, $Z(\rightarrow
l^+l^-)Z(\rightarrow q\bar q)$, $W(\rightarrow l\nu)Z(\rightarrow
q\bar q)$, $W(\rightarrow l\nu)W(\rightarrow q\bar q)$,
$Z(\rightarrow l^+l^-) + jets$, $W(\rightarrow l\nu) + jets$ and
$W(\rightarrow l\nu)q\bar q$ \cite{Marchesini:1992ch,Mangano:1993kp}.
At the Tevatron, the selection criteria require at least four jets in
each event, which must satisfy $E_T> 15$ GeV and $|\eta|<20$.

\begin{figure}[ht]
 \centerline{
    \includegraphics[width=0.44\textwidth,clip]{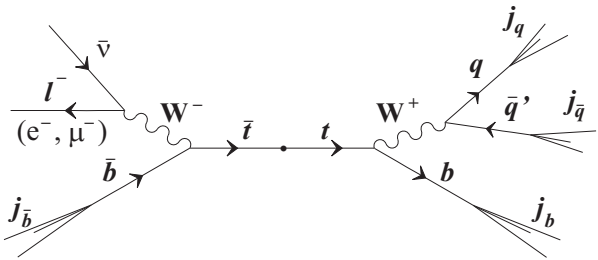}}
  \caption{Schematic example of the top and anti-top decays in the lepton plus jets channel.}
  \label{fig:avto1}
\end{figure}

\begin{figure}[ht]
    \begin{center}
        \includegraphics[width=0.35\textwidth,clip]{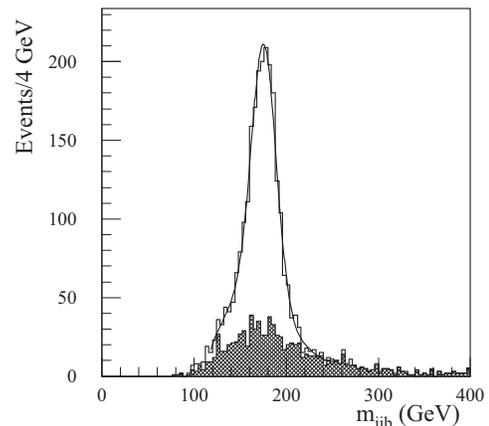}
    \end{center}
\vskip -0.8cm
    \caption{Invariant mass \mjjb~distributions from full simulation
    for the LHC for the lepton plus jets channel in top anti-top production
    \cite{mtdetLHC}.}
    \label{lhcmjjb}
\end{figure}

 There are twenty-four distinct ways to assign
these four jets to the four partons $q$, $\bar q'$, $b$ and $\bar b$.
The CDF and D0 experiments
 select the combination with a minimum value for $\chi^2$ in
their statistical analysis of the process \cite{CDFl+jets,D0l+jets}.
Furthermore, the two jets with an invariant mass closest to \mW~can
be associated to the $W\rightarrow q\bar q$ decay, and the $j_q
j_{\bar q} j_{b}$ or $j_q j_{\bar q} j_{\bar b}$~ combination having
the highest \pt~can be associated to the top or anti-top decay
~\cite{atlasphysnote99024}. With sufficient resolution, therefore,
we should find two peaks (if $m_t$ and $m_{\bar t}$ are different)
when we plot the events as a function of the invariant mass \mjjb~
(See Figure \ref{lhcmjjb}).

The CDF and D0 experiments
find consistent results with the assumption that $m_t=m_{\bar t}$.
Their analysis uses the relations:
\begin{eqnarray}
\label{eq:constraints}
m(t\rightarrow l^+\nu b) &=& m(\bar t\rightarrow q \bar q'\,\bar b)\nonumber\\
m(l\nu) &=& m_W\\
m(q\bar q) &=& m_W.\nonumber
\end{eqnarray}
(We know that the $W^+,W^-$ masses are very similar from other
experiments (see Table \ref{$CPT$}), but a more detailed analysis of
$R_{CPT}(t)$ should eliminate the first of the above equalities.)

We can now find a bound on the absolute value of $R_{CPT}$
in a manner similar to the previous section. In this case, we find
a stronger constraint, as shown in Table \ref{l+jets}.  The constraints
are dominated by the systematic errors due to our conservative
approach. The major contributions to these errors come from the
uncertainties in the jet energy scale and in gluons with high
transverse momentum \cite{CDFl+jets,D0l+jets}.

\begin{table}[ht]
\centering
\begin{tabular}{||c|c||}
\hline\hline
Experiment& $|R_{CPT}(t)|$ (95\% c.l.)
\\
\hline
D0\cite{D0l+jets}&$<\,10.9\times 10^{-2}$
\\
CDF\cite{CDFl+jets}&$<\,10.1\times 10^{-2}$
\\
\hline\hline
\end{tabular}
\caption{\label{l+jets} Estimates of the present experimental
constraints on the absolute value of the difference mass ratio
between top ($t$) and anti-top
 ($\bar{t}$): $R_{CPT}(t)\equiv 2(m_t-m_{\bar t})/(m_t+m_{\bar t})$. The numbers listed
 above are upper bounds from the lepton plus jets channel on this ratio at 95\% c.l.}
\end{table}

The systematic errors on the top and anti-top mass measurement will
be reduced at the LHC down to $1$ GeV by using top and anti-top
samples with high \pt  \cite{atlasphystdr,atl:iefthy99}.
 The special topology of
 such a sample reduces the combinatorial errors and
backgrounds. \ATLAS~performed a preliminary study of this possibility
reconstructing \mt~ or $m_{\bar t}$~ from the three jets in the one
hemisphere (\mt=\mjjb~ or $m_{\bar t}=m_{jjb}$) and by summing up the
energies in the calorimeter towers in a
large cone around the top direction. The analysis is similar to the inclusive case; 
a mass difference will show up as a double peak in the mass
reconstruction.

A typical sample of $10^4$ events for an integrated luminosity of
10~\infb could be collected leading to a statistical error of
$\pm$0.25~GeV, which is well below the systematic uncertainty (see
e.g. \cite{mtdetLHC}, where \mt$=m_{\bar t}$~ is supposed). Repeating
the analysis of the previous section, the expected sensitivity is
found to be $|R_{CPT}(t)|\simeq 0.014$ at 95\% c.l. or equivalently,
$m_t-m_{\bar t}\simeq 2.4$ GeV.

\section{Prospects for the Linear collider}

At the Intenational Linear Collider,
the top quark mass can be determined with a
precision of the order of 100 MeV with only 11 fb$^{-1}$ of data
using threshold scan analyses \cite{threshold}.
This mass is determined directly by the accelerator energy at which
one sees the onset of $t \bar t$ production. However, this study
cannot be used to determine a mass difference between top and
anti-top, since the threshold depends on the sum of both masses.

However, at the ILC, it is possible to study the same analyses that
we have  performed for hadronic colliders, i.e. the lepton plus jets
and the di-lepton channel. There are fewer studies about the
determination of the top mass in these channels since they have less
precision than the threshold scan. 
However the collision energy is smaller, 
this collider has several advantages such as its cleaner experimental
environment, the possibility of polarized beams and tunable
collision energies \cite{ILCLHC}. For these reasons, 
we expect an increase in the statistical uncertainties but a decrease
of the systematic ones 
\cite{Biernacik:2003xv}. The systematic errors are most important (at
least in our conservative approach) and this leads to a small
improvement of the sensitivity in relation to the LHC, but always at
the order of 1 per cent for both channels.

\section{Conclusions}

We will learn a great deal about particle physics in the next few
years.  We can  expect several surprises at the LHC and at the
proposed International Linear Collider.  Furthermore, precision
measurements inside the top sector are just beginning with the second
run at the Tevatron.

In this note, we have studied the possibility that these new
experiments will observe 
$CPT$ violation in the top sector. We have analyzed the viability of
a measurement in the mass difference between top and anti-top,
which we have parameterized through 
$R_{CPT}(t)\equiv 2(m_t-m_{\bar t})/(m_t+m_{\bar t})$. We have
estimated the current constraints on this parameter from Tevatron Run
I data. This analysis shows constraints around 10\% coming mainly from the
lepton plus jets channel, as shown in Table \ref{combo}.

\begin{table}[ht]
\centering
\begin{tabular}{||c|c||}
\hline\hline
Process& $|R_{CPT}(t)|$ (95\% c.l.)
\\
\hline
$l+l$&$<\,13\times 10^{-2}$
\\
$l+jets$&$<\,10\times 10^{-2}$
\\
\hline\hline
\end{tabular}
\caption{\label{combo} Estimates of the present experimental
constraints on the absolute value of the difference mass ratio
between top ($t$) and anti-top
 ($\bar{t}$): $R_{CPT}(t)\equiv 2(m_t-m_{\bar t})/(m_t+m_{\bar t})$. The numbers listed
 above are upper bounds on this ratio at 95\% c.l from the data collected at Tevatron in the top anti-top production.}
\end{table}

We have also estimated the sensitivity of the LHC
to $R_{CPT}(t)$, and have shown that the constraints can be improved
by an order of magnitude over current values.
The ILC would be
able to improve the LHC precision but not as much as for the
determination of the sum of top and anti-top masses.

A challenge for these future collider experiments is to check $CPT$
invariance in the top sector with an accuracy of $2\%$ or
better. For $R_{CPT}$ 
we have argued that this objective could be achievable if all sources
of uncertainty are kept to the order of $1\%$ level. A more detailed
analysis and estimation of the systematic uncertainties is in
progress.

\vspace{.4 cm} {\bf Acknowledgments:} We thank J. L. Feng, A. Juste, Y. Kiyo and
F. Petriello for their comments and suggestions. The work of JARC is
supported in part by NSF grant No.~PHY--0239817, the Fulbright-MEC
program, and the FPA 2005-02327 project (DGICYT, Spain). The work of
AR is supported in part by NSF Grant No.~PHY--0354993.  The work of
FT is supported by the NSF grant No.~PHY-0355005.

\end{document}